\documentclass[twoside,11pt]{article}
\usepackage{jair, theapa, rawfonts}

\usepackage[utf8]{inputenc} 
\usepackage[T1]{fontenc}    
\usepackage{url}            
\usepackage{amsfonts}       
\usepackage{nicefrac}       
\usepackage{microtype}      
\usepackage{xcolor}         
\usepackage{graphicx}
\usepackage[font=small]{caption}
\usepackage{textcomp}

\newcommand{\textapprox}{\raisebox{0.5ex}{\texttildelow}}

\begin{document}
\title{Over-communicate no more: \\ Situated RL agents learn concise communication protocols}

\author{\name Aleksandra Kalinowska \email ola@u.northwestern.edu \\
       \name Elnaz Davoodi \email elnazd@deepmind.com \\
       \name Florian Strub \email fstrub@deepmind.com \\
       \name Kory W.~Mathewson \email korymath@deepmind.com \\
       \name Ivana Kajić \email kivana@deepmind.com \\
       \name Michael Bowling \email bowlingm@deepmind.com \\
       \name Todd D.~Murphey \email t-murphey@northwestern.edu \\
       \name Patrick M.~Pilarski \email ppilarski@deepmind.com}

\maketitle

\begin{abstract}

While it is known that communication facilitates cooperation in multi-agent settings, it is unclear how to design artificial agents that can learn to effectively and efficiently communicate with each other. 
Much research on communication emergence uses reinforcement learning (RL) and explores \textit{unsituated} communication in one-step referential tasks---the tasks are not temporally interactive and lack time pressures typically present in natural communication. 
In these settings, agents may successfully learn to communicate, but they do not learn to exchange information concisely---they tend towards over-communication and an inefficient encoding. 
Here, we explore \textit{situated} communication in a multi-step task, where the acting agent has to forgo an environmental action to communicate. Thus, we impose an opportunity cost on communication and mimic the real-world pressure of passing time. We compare communication emergence under this pressure against learning to communicate with a cost on articulation effort, implemented as a per-message penalty (fixed and progressively increasing). 
We find that while all tested pressures can disincentivise over-communication, \textit{situated} communication does it most effectively and, unlike the cost on effort, does not negatively impact emergence.
Implementing an opportunity cost on communication in a temporally extended environment is a step towards embodiment, and might be a pre-condition for incentivising efficient, human-like communication.
\end{abstract}

\section{Introduction}
\label{Introduction}

Effective communication is a key skill for collaboration in a multi-agent setting~\cite{chopra2020evaluation}. As humans, we share communication protocols and cooperative conventions that have evolved over thousands of generations to optimize communication efficiency. As an example, we communicate in accordance with cooperative principles, such as Grice's maxims of conversation~\cite{grice1975logic}. In line with the maxim of quantity, we are known to try to be as informative as possible, giving only as much information as is needed~\cite{grice1975logic}. If future artificial systems are to cooperate with humans, it will be beneficial for their communication protocols to follow these patterns~\cite{crandall2018cooperating,steels2003evolving}. As a result, understanding the process of communication emergence and the pressures that shape the emergent communication protocols is of interest to the scientific community.

With a recent increase in available computational power, the field has seen a lot of progress with communication successfully emerging between reinforcement learning (RL) agents in a range of learning environments~\cite{wagner2003,angeliki-review}.  However, prior work shows that the emerged communication protocols often do not share properties of natural languages~\cite{kottur2017natural} and that artificial agents tend towards an anti-efficient encoding~\cite{anti-efficient-rahma}. This likely happens because in many of the studied environments, communication is not \textit{situated}---the action space of the agent does not include both communicative and environmental actions~\cite{wagner2003,crawford1982strategic}. Agents do not learn to reason about \textit{whether} to communicate; instead, communication is guaranteed and free to the agents. When free, excessive use of communication does not negatively affect the outcome of the game (or cause agent frustration), as it might in a real-world situation~\cite{steels1995artificial}. As a possible solution, we explore \textit{situated} communication and show it is possible to obtain concise communication protocols by providing the agent with an action-communication trade-off.

In this work, we compare pressures that can incentivise conciseness during communication emergence. As our testbed, we use a cooperative multi-step navigation task with two RL agents. In the task, a speaker provides navigation hints to help a listener reach a goal within a gridworld maze. We explore three training regimes: 
(i) \textit{unsituated} communication (cheap talk); the speaker sends a message to the listener at each timestep without any cost, similar to the communication paradigm in existing work~\cite{anti-efficient-rahma,ease-of-teaching,lazaridou2018emergence},
(ii) \textit{unsituated} communication with a per-message penalty; the speaker experiences a cost on communication effort (either fixed or progressively increasing), similar to prior work~\cite{lazimpa} when fixed, 
(iii) \textit{situated} communication as introduced by Wagner et al. (2003); the listener has to forgo an action to solicit information from the advising agent, experiencing an opportunity cost on communication. 
Using the collaborative navigation task, we evaluate how the different pressures in the three training regimes can incentivise sparse information sharing during communication emergence. We find that \textit{situated} communication (regime iii) outperforms an internal cost on communication effort (regime ii) in terms of both conciseness of the emerged communication protocol and overall task performance. 

\section{Background}
\label{background}
Initially, emergent communication between RL agents was largely studied in one-step referential games, such as the Lewis task~\cite{anti-efficient-rahma,ease-of-teaching,lazaridou2018emergence,kajic2020,choi2018compositional}. The Lewis task~\cite{Lewis1969-LEWCAP-4} is a cooperative game, where the speaker sees an artifact (e.g., an image) and communicates a message from a fixed vocabulary (e.g., a symbol) to the listener, who then interprets the message to select a target item from among a set of distractors. This type of learning environment is known to successfully enable language development~\cite{kirby2002emergence}. However, this setting does not recreate the temporal aspects of a real-world environment, which may influence the structure of the emergent languages.
 
Recent work increasingly explores multi-step tasks that enable temporally extended dialogue, more similar to real-world environments~\cite{mordatch2018emergence,cao2018emergent,evtimova2018emergent,bouchacourt2019miss,eccles2019biases,das2019tarmac,jaques2019social}. Cao et al. (2018) investigate multi-step negotiation. Jaques et al. (2019) test communication emergence during Sequential Social Dilemmas in environments, such as Harvest or Cleanup. Evtimova et al. (2018) propose a multi-modal, multi-step referential task (a modified version of the Lewis task). Bouchacourt and Baroni (2019) introduce a fruit-tool matching game, similar to a multi-step Lewis task but with a preference-based reward. Here, we build on this body of work by introducing a multi-step navigation task in a gridworld environment. In Appendix~\ref{appendix: multistep-vs-upfront}, we elaborate on how our task compares to existing work and on the direct impact of introducing multi-step communication in this task. 

The literature predominantly considers two types of communication: (i) communication through a cheap talk channel~\cite{crawford1982strategic,farrell1996cheap} and (2) communication through environmental actions~\cite{shoham2008multiagent}. In cheap talk, agents have a designated communication channel where they share messages at every timestep---the messages are free to the agents and agents do not reason about \textit{whether} to communicate. Many authors find that selfish agents do not learn to effectively use an ungrounded cheap talk communication channel~\cite{cao2018emergent,choi2018compositional}. However, one can incentivise communication in a cheap talk channel by rewarding an intrinsic motivation to influence others~\cite{jaques2019social} or by introducing a pro-social reward~\cite{foerster2016learning,lazaridou2018emergence}, as we do in our work. On the other hand, when communicating through actions, agents' environmental actions are made visible to other agents, enabling the environmental actions to be used for communication. Note that environmental actions are binding---they impact the state of the environment and the resulting reward, so agents do not have an inconsequential way to exchange information as with a cheap talk channel. In these scenarios, cooperation has a high success rate~\cite{cao2018emergent,noukhovitch2021emergent,carroll2019utility}, but the communication is inflexible and directly connected with the task---there is mixed opinion whether agents are actually communicating~\cite{lowe2019pitfalls}. In this work, we evaluate a third type of communication that we refer to as \textit{situated} communication as introduced by Wagner et al. (2003) and used e.g. by Lowe et al. (2019) in matrix games, such as the Prisoner's Dilemma. When using \textit{situated} communication, the listener has to actively choose between communicative and environmental actions. While only environmental actions directly affect the environment and allow the agent to obtain a reward, the agent can choose to forgo an environmental action to communicate.

In most studies, the emerged language structures are analyzed for shared commonalities with natural languages, such as compositionality or encoding efficiency. Although desired, it is nontrivial for such properties to emerge spontaneously~\cite{kottur2017natural}. As a result, researchers introduce pressures during training and/or structure the learning process to incentivise specific language properties. To encourage compositionality in the emergent communication protocol, one can introduce populations of agents~\cite{chaabouni2021emergent,rita2022on} or the need for a language to be easily teachable~\cite{ease-of-teaching}. Efficient communication can be incentivised by modifying the agents' reward structure, e.g., by adding an internal cost of articulation~\cite{lazimpa} or by prioritizing messages based on a metric of confidence~\cite{zhang2019efficient}. Sparse communication has also been shown to emerge in mixed and competitive settings, where communicating too much might harm agent performance, e.g., in the case of prey coordinating an escape from predators~\cite{singh2018learning}. In parallel, researchers have studied communication patterns during competitive games---e.g., auctions---where the capacity of the communication channel is externally bounded~\cite{blumrosen2007auctions}. Here, we propose \textit{situating} the communication in a multi-step task as a mechanism to shape the properties of the emergent communication protocol. 

In our experiments, we study the emergence of communication in a cooperative multi-step navigation task. Like humans or robots that can only observe a small part of the world in their proximity, the listener has a limited view of its environment and has to rely on the speaker for guidance~\cite{denis1999spatial}. The task is an abstraction of a real-world task~\cite{de2018talk} where a person receives guidance from an oracle while navigating towards a goal. As our primary testbed, we consider a maze with T-junctions (i.e. junctions with a left/right turn)~\cite{deacon2006t}. The well-defined decision points enable us to more easily quantify the amount of information exchanged by the agents. 

\section{Environment}
\label{sec: environment}

\begin{figure*}[t]
\begin{center}
\includegraphics[width=\textwidth]{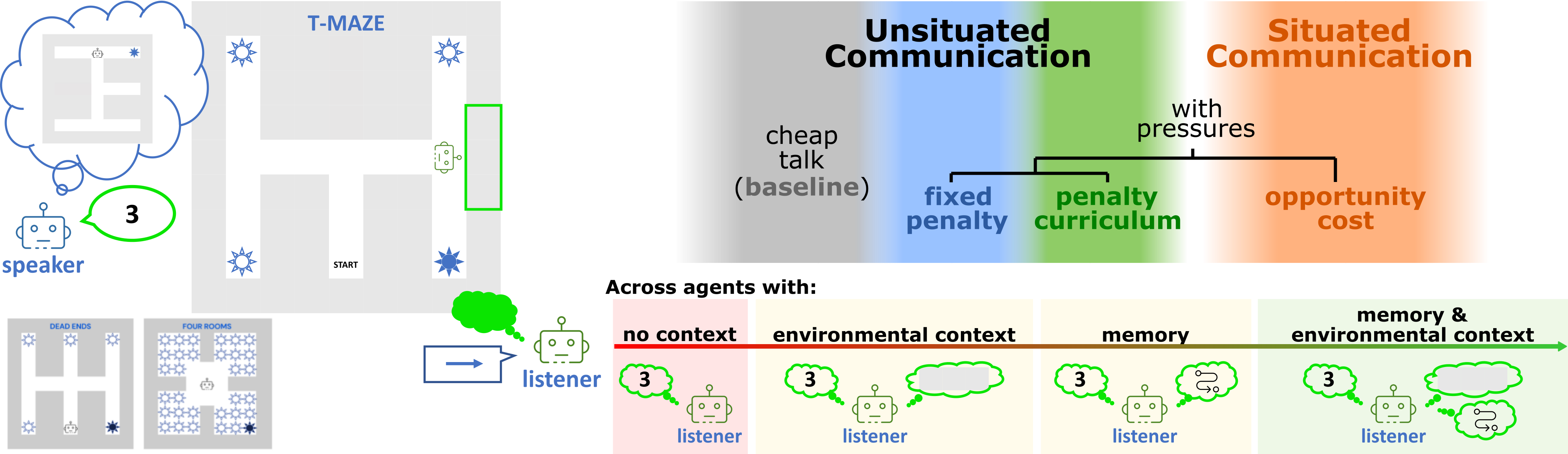}
\end{center}
\caption{\textbf{Experimental setup.} (left) Gridworld environment \& navigation task, (right) experimental conditions.} 
\label{fig: task}
\end{figure*}

We define a cooperative navigation task as a Markov Decision Process (MDP) with two RL agents. We test three environments: T-maze, dead ends, and four rooms, as shown in Fig.~\ref{fig: task}. The T-maze serves as our primary testbed. All three environments are set up as a pixel-based gridworld with a maze inside. Features of the world are encoded with binary vectors and represented with colors: walls are black ($[1,1,1]$), the maze is white ($[0,0,0]$), the agent is green ($[0,1,0]$), and the target is blue ($[0,0,1]$), as shown in Fig.~\ref{fig: active-listener-walkthrough} in the Appendix. The full environmental representation is defined as a $9\times9\times3$ matrix (height by width by RGB).

\textbf{The agents.} There are two independent RL agents, a speaker and a listener (i.e. acting agent). The speaker does not reside within the gridworld and cannot take environmental actions (i.e. navigate the maze) but instead can communicate information to the listener. The speaker's action (i.e. message) space spans $5$ symbols $[0, 1, ..., 4]$. We refer to the message $m_t = 0$ as a null message, and $m_t \in [1, 2, 3, 4]$ as non-zero messages. At each timestep, the speaker can see the entire gridworld, including the location of the agent and the location of the goal. The speaker's view of the world map is rotated to align with the direction that the listener is facing. The listener is embedded in the gridworld and can take actions to move through the maze. The action space of the listener spans $5$ actions [move up, move down, move right, move left, stay in place]. The listener's observation consists of the environmental view (if any) concatenated with the message from the speaker. We test the listener under two conditions: (1) with no visibility, where the listener's observation consists solely of the speaker's message, and (2) with partial visibility, where the listener can see the $3$ pixels directly in front of them. The second variant gives the listener some environmental context to take actions without needing to rely solely on communication. We test agents with and without memory. Agents without memory have to rely only on their current observations to take an action. Agents with memory have an internal representation of the history of an episode---they can use accumulated knowledge from prior timesteps to make decisions in the current timestep.  

\textbf{The task.} The goal of the agents is to cooperate so that the listener reaches the target. In each experimental episode, both agents receive a reward $R=1$ if the listener reaches the target before the episode terminates. Episode timeout is set to $100$ steps with a gamma discount factor of $\gamma=0.99$. In all three environments, the goal locations are randomly assigned to one of the pixels indicated with a star in Fig.~\ref{fig: task}. There are $4$, $5$, and $32$ possible goal locations in the T-maze, dead ends, and four rooms environments, respectively. In each episode, the listener agent starts from the `START' cell indicated in Fig.~\ref{fig: task}. In the T-maze, the agent always starts from the bottom middle cell and the goal locations are randomly assigned to one of the $4$ corners of the maze. 

\section{Situated vs. Unsituated Communication}
\label{sec: communication modes}

Wagner et al. (2003) classify simulated environments for communication emergence based on the action space of the simulated agents. They define two types of actions: (1) communicative (i.e. sending or receiving signals)---actions that do not affect the state of the world or other agents, and (2) non-communicative (i.e. environmental)---actions that affect the environment and/or modify the agent's own internal state. Depending on the actions available to the agents, they define \textit{situated} and \textit{unsituated} simulations of communication emergence:
\begin{enumerate}
    \item \textit{unsituated}: an agent's actions are only communicative (or only non-communicative)
    \item \textit{situated}: an agent's actions include \textit{both} communicative and non-communicative actions.
\end{enumerate} 
Given these definitions, we define two modes of communication based on the actions available to the listener. During \textit{unsituated} communication in our task, the speaker generates a $1$-token message at every timestep and the message is broadcasted to the listener before they choose an action. We refer to the listener using this mode of communication as a passive listener, because it passively receives the speaker's message at every timestep. When messages are free, this communication mode is equivalent to cheap talk. In \textit{situated} communication, the listener can actively choose between (i) taking an environmental action and (ii) soliciting a message. The message is only broadcasted to the listener after they ask for information---we refer to this listener as active. The active listener can solicit to receive information in a following timestep by choosing a \textit{stay in place} action at the current timestep. The active listener experiences an opportunity cost to communication---they have to forego an environmental move (that could bring them closer to the target) to obtain information and make an informed decision. As a result, they have to learn \textit{whether} to communicate at all. 

Similar to much of the existing work in the field~\cite{angeliki-review}, the communication in our experiments is asymmetrical (i.e. the action space of the speaker and listener are different). During \textit{unsituated} communication, we simulate a passive listener that can take environmental actions and receive signals from a speaker but cannot send signals itself. If the agents' action spaces were symmetrical (the scenario considered by Wagner et al., 2003), this would be a somewhat degenerate case---communication would never emerge between two agents who can only receive signals and take environmental actions. As a result, our definition of \textit{unsituated} communication deviates slightly from the definition introduced by Wagner et al. (2003). 

\section{Experimental Setup}
\label{sec: experimental setup}

\textbf{Agent architectures.} The speaker and the listener are designed as two independent RL agents. Both agents have the same architecture without sharing weights or gradient values. They both have a 2-layer Convolutional Neural Network (CNN) that generates an $8-32$ bit representation $s$ of the environment. In the case of the listener, this representation of the environment gets concatenated with the message received from the speaker. In both cases, the vector gets passed into a fully connected layer that generates the agent's action. Agents with memory, have an additional single-layer LSTM~\cite{lstm} after their fully connected layer. 

We train the agents using neural fitted Q~learning~\cite{riedmiller2005neural}, with an Adam optimizer~\cite{adam-opt} and $Q_t(\lambda$) where $\lambda =0.9$. The Q values are updated using temporal difference (TD) error where the bootstrapped $Q_t(\lambda$) is defined as follows:
$$
    Q_t(\lambda) = (1-\lambda) \sum_{n=1}^{\infty} \lambda^{n-1} Q_t^{(n)}
$$
During training, agents use an $\epsilon$-greedy policy with the exploration rate set as $\epsilon=0.01$.

\textbf{Hyperparameters.} For each experiment, we run a hyperparameter sweep over learning rates of the speaker and listener $\alpha \in \{10^{-5}, 10^{-6}\}$ and over the size of the environmental representation $s \in \{8, 16\}$, which denotes the size of the output vector of the convolutional layers of the agents' networks. We run the simulation with each hyperparameter setting $10$ times with different random seeds. For each experiment, we present the best mean over the $10$ replicas. The best mean is selected based on the metric of solution optimality (the normalized reward per step). When we plot metric means, we include the standard error of the mean.

\textbf{Message penalties.} In \textit{unsituated} communication, we test the consequences of introducing a cost of articulation for the speaker. In the baseline scenario (i.e. cheap talk), all messages are free to the agent. With a cost of articulation, each non-zero message incurs a penalty, while a null message (the symbol $0$) remains free to the agent. We test two penalty mechanisms: a fixed per-message penalty and a progressive per-message penalty. When fixed, the per-message penalty remains constant throughout the entire duration of training. We test fixed penalties with values $m_p \in \{0.01, 0.05, 0.1\}$. 

As the second mechanism, we introduce a penalty curriculum with a scheduler that progressively increases the penalty between stages. The penalty scheduler is a mapping from the curriculum stage to the penalty in that stage. We test two curriculum implementations with the following mappings $m_{p1}=\{0: 0, 1: 0.01, 2: 0.02, 3: 0.03, 4: 0.04, 5: 0.05, ...\}$ and $m_{p1}=\{0: 0, 1: 0.01, 2: 0.05, 3: 0.1, 4: 0.2, 5: 0.3\}$.
Both curricula start with no per-message penalty in stage $0$, allowing cost-less exploration of a successful communication protocol. In the next stage, both curricula impose a small penalty that progressively increases over the time of training. Intuitively, the first curriculum is more gradual, enabling a more stable solution; the second curriculum is more rapid, possibly speeding up the learning process. The agents can progress to the next stage of the curriculum after $2M$ or $5M$ training steps if they achieve a performance threshold or if they spend $15M$ steps in their current stage. We test 3 progression thresholds based on the agents' success rate in reaching the target: $92\%$, $95\%$, and $97\%$.  

\begin{figure}[t]
\begin{center}
\includegraphics[width=\textwidth]{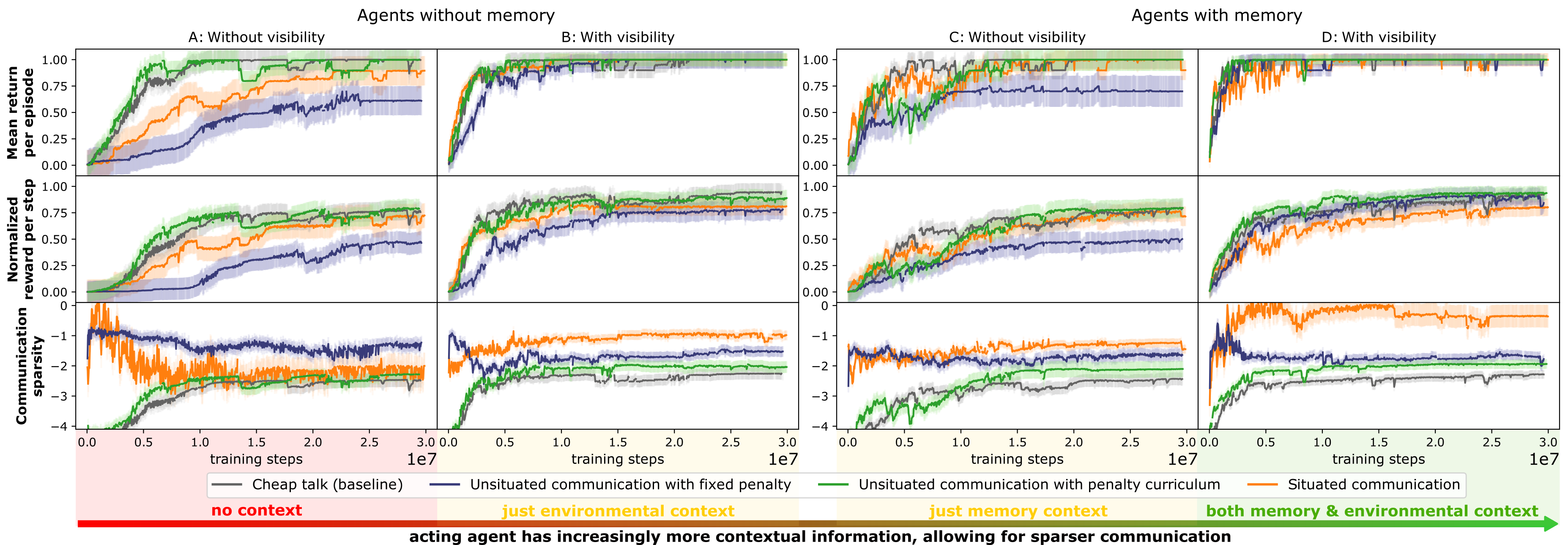}
\end{center}
\caption{\textbf{Comparison of pressures for avoiding over-communication (mean over 10 seeds) in the T-maze.} Top row illustrates an average metric of success---under almost all conditions agents are able to find a solution to the task. Middle row illustrates an average metric of optimality---with context (i.e., visibility and/or memory) agents require fewer steps to reach the target. Bottom row illustrates communication conciseness---\textit{situated} agents are able to communicate most sparingly. }
\label{fig: pressures comparison}
\end{figure}

\textbf{Evaluation metrics.} We evaluate agent performance according to three criteria: (i) task success (via the mean return per episode), (ii) solution optimality (via the normalized reward per step), and (iii) communication efficiency (via communication sparsity). The metric of task success is calculated after each episode as 
\begin{equation}
M_{t}(n) = (\sum_{i=1}^n R_{i})/n,
\end{equation}
where $n$ is the number of episodes thus far and $R_i$ is the reward for the $i$th episode. The metric represents the likelihood of the agents succeeding at reaching the target before episode timeout. When it converges to $1$, agents are reliably reaching the target in each episode. The normalized reward per step quantifies the optimality of the path taken to solve the task. If the task is solved in the optimal number of steps ($s_{opt} = 9$), agents obtain a per-step reward of $1$. Formally, it is calculated as 
\begin{equation}
M_{o}(n) = s_{opt}\sum_{i=1}^n R_{i}/s_{i},
\end{equation}
where $s_i$ is the number of steps the agent took in episode $i$. Finally, the metric of communication sparsity quantifies the efficiency of information exchanged between the collaborating agents. We define it as the average negative logarithm of the number of non-zero messages generated by the speaker per episode, such that 
\begin{equation}
M_{s}(n) = (\sum_{i=1}^n -\log m_{i})/n,
\end{equation}
where $m_i$ is the number of non-zero messages in episode $i$. Agent pairs that converge to a higher value form a more efficient communication protocol---they exchange fewer messages per episode. If agents were able to solve the task using a single message, their sparsity metric would equal $0$. As an example, in the T-maze, depending on the listener's characteristics: partial or no visibility, the optimal number of messages is equal to two ($M_s=-0.7$) or nine ($M_s=-2.2$) messages per episode for agents without memory, and one ($M_s=0$) message per episode for agents with memory.

\section{Results}

We experimentally compare two pressures that can incentivise a concise exchange of information during communication emergence. Specifically, we consider two types of pressure:

\begin{itemize}
    \item A per-message penalty: similar to a cost of articulation. During \textit{unsituated} communication, we impose a cost on effort through a penalty on non-zero messages. We test (a) a fixed penalty, and (b) a penalty curriculum with a progressively increasing per-message penalty.
    \item An opportunity cost on communication: mimicking the pressure of time in a real-world situation. We study agents using \textit{situated} communication, where at each step the listener faces an action-communication trade-off. 
\end{itemize}

In our experiments, we train agent pairs communicating (1) using cheap talk (baseline), (2) using \textit{unsituated} communication with a fixed penalty or a penalty curriculum, and (3) using \textit{situated} communication. In all experiments, we train agents under two visibility and two memory conditions (see Sections~\ref{sec: environment} \&~\ref{sec: experimental setup} for more details). Without visibility or memory, the acting agent relies solely on the speaker's most recent message to take an action. With visibility and memory, the acting agent has increasingly more contextual information, allowing for sparser and more efficient communication.

\begin{figure*}[t]
\begin{center}
\includegraphics[width=\textwidth]{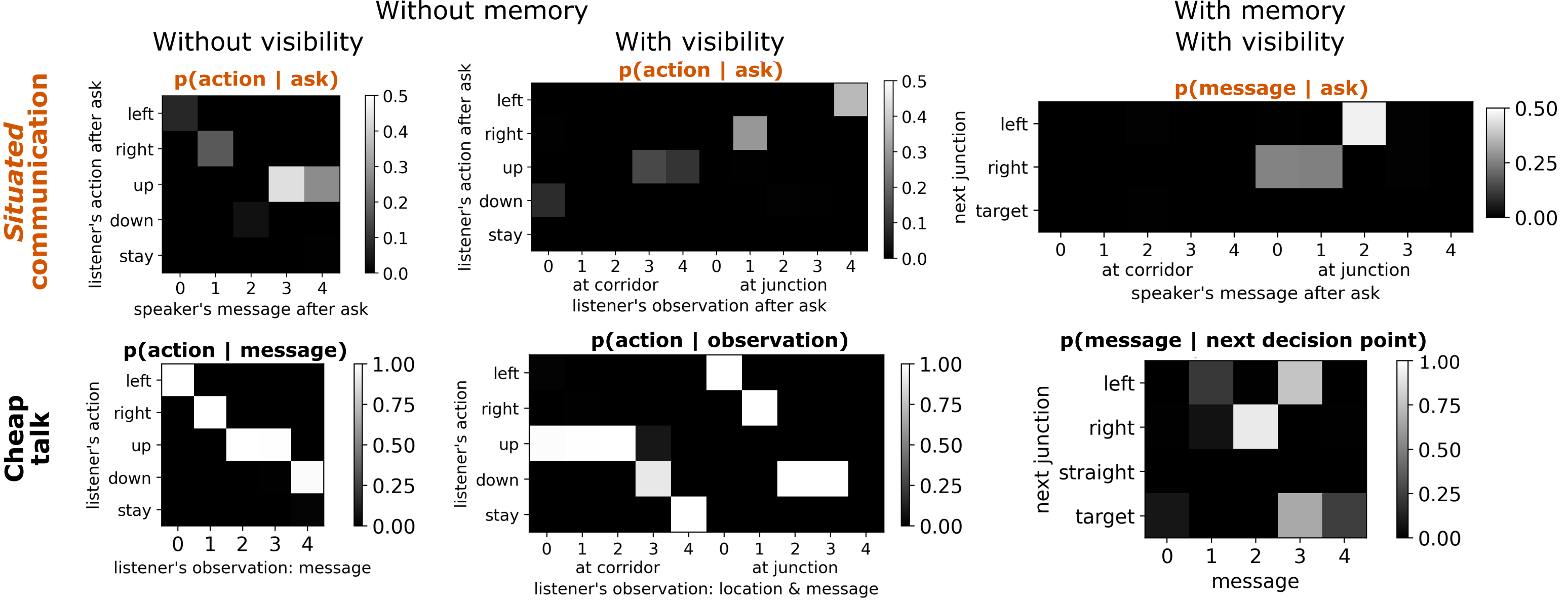}
\end{center}
\caption{\textbf{Communication protocols for successful agent pairs in a T-maze environment.} Communication protocols are analogous when the listener has no context (left-most column). With context, agents with \textit{situated} communication learn to exchange information sparsely, communicating mostly at the junctions. } 
\label{fig: heatmaps}
\end{figure*}

\subsection{Cheap talk (baseline)} 

\textbf{Without pressures to be concise, agents successfully learn to solve the task.} We start by analyzing the learning pattern and behavior of agents without communication constraints---we use their performance as a reference in our subsequent analysis. When allowed to communicate, all agents in the T-maze environment (grey lines in Fig.~\ref{fig: pressures comparison}) learn to solve the task---their mean return per episode converges to $1$. Best agent pairs find an optimal solution---their normalized reward per step converges to $1$. With both memory and partial visibility for context, even the average performance is close to optimal (refer to Fig~\ref{fig: pressures comparison} column D, row 2). Moreover, we analyze the sparsity of the established communication protocol. With \textit{unsituated} communication and no pressures (cheap talk), agents have no incentive to be sparse. As a result, for all tested conditions, their sparsity metric is equal to the number of steps needed to solve the task, and lowest compared to agents with added pressures (refer to the grey lines in Fig.~\ref{fig: pressures comparison} row 3).

For the best-performing agent pairs, we analyze the communication protocol qualitatively as illustrated in Fig.~\ref{fig: heatmaps}. Best-performing agent pairs agree on unambiguous meanings of messages and, in some cases, learn synonyms to signal the same environmental action. Under partial visibility (refer to the bottom middle heatmap in Fig.~\ref{fig: heatmaps}), the meaning of messages depends on the environmental context (e.g. message $1$ at the corridor might be consistently interpreted by the listener as `move up' but at the junction as `move right'). All else equal, successful agents without memory converge to a just-in-time protocol, where at each time step the listener can unambiguously interpret the speaker's message. Interestingly, memory influences timing in the established communication protocols. The best-performing agents with memory learn a look-ahead communication protocol. As an example, a speaker with memory might broadcast the same message for the first $4$ steps of a T-maze episode, alerting the listener to make a left or right turn at the junction (see the bottom right heatmap in Fig.~\ref{fig: heatmaps}).

\begin{figure*}[t]
\begin{center}
\includegraphics[width=\textwidth]{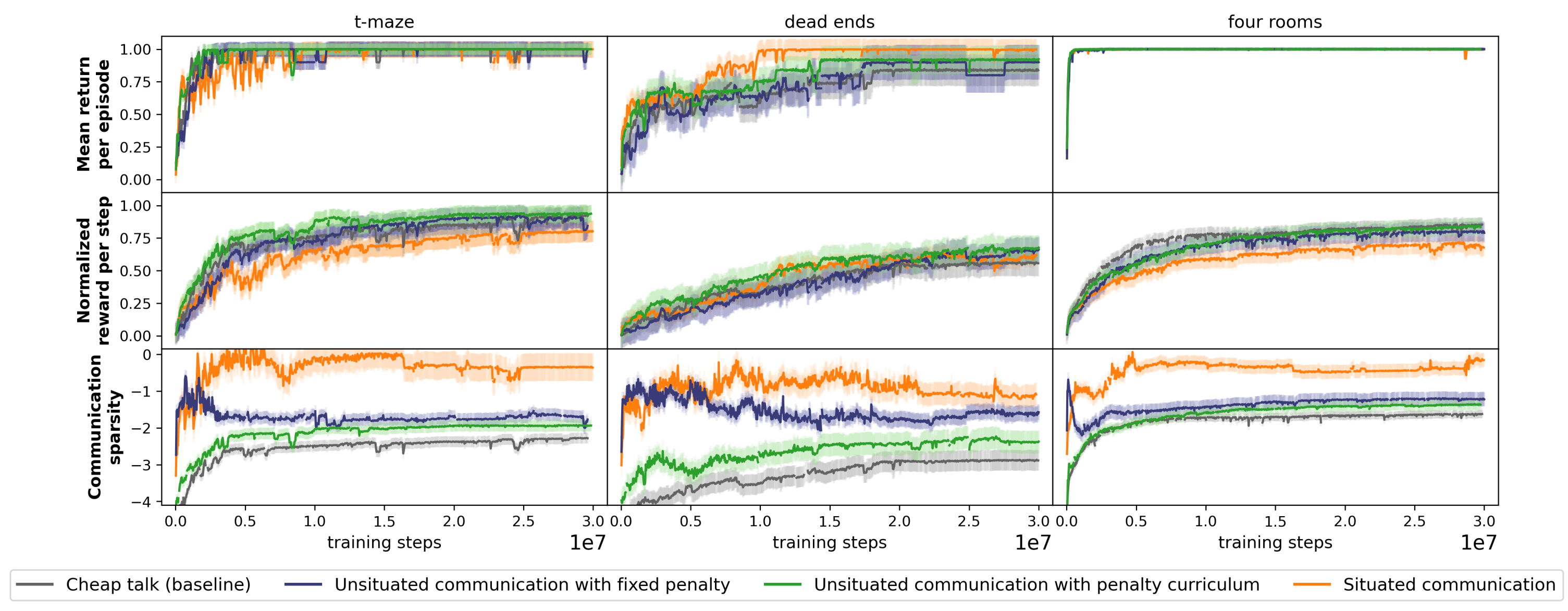}
\end{center}
\caption{\textbf{Comparison of pressures for avoiding over-communication (mean over 10 seeds) in the 3 environments.} All agents have memory; listener has partial visibility. Across all 3 environments, situated agents agree on the most concise communication protocols without sacrificing their ability to solve the task. } 
\label{fig: comparison 3env}
\end{figure*}

Lastly, we analyze agent performance in two other environments. The learning curves are illustrated in Fig.~\ref{fig: comparison 3env}. In four rooms, baseline agents achieve similar performance to the T-maze environment. The dead ends maze is more difficult---not all agents are able to solve the task. Importantly, compared to the T-maze, both dead ends and four rooms are significantly different from the listener perspective. In dead ends, when the listener sees a corridor (i.e. wall-path-wall alternating pixels), it can no longer learn to always go straight---$4$ out of $5$ times it has to turn when seeing a corridor at the first junction. As a result, it has to rely more on the messages from the speaker. Similarly, in four rooms, there are many possible goal locations ($32$ compared to $4$ in the T-maze) and the environmental context is ambiguous---visibility does not help the listener with decision-making as much as it does in the T-maze. Again, the listener has to rely more on the messages. Because the dead ends and four rooms environments are more ambiguous than the T-maze environment from the listener perspective, we only consider listeners with context (i.e. memory and partial visibility). With added pressures, this allows the agent pair to learn to communicate more sparingly and still succeed at the task. 

\subsection{\textit{Unsituated} communication with a penalty (fixed and progressive)}

\begin{figure}[t]
\begin{center}
\includegraphics[width=\textwidth]{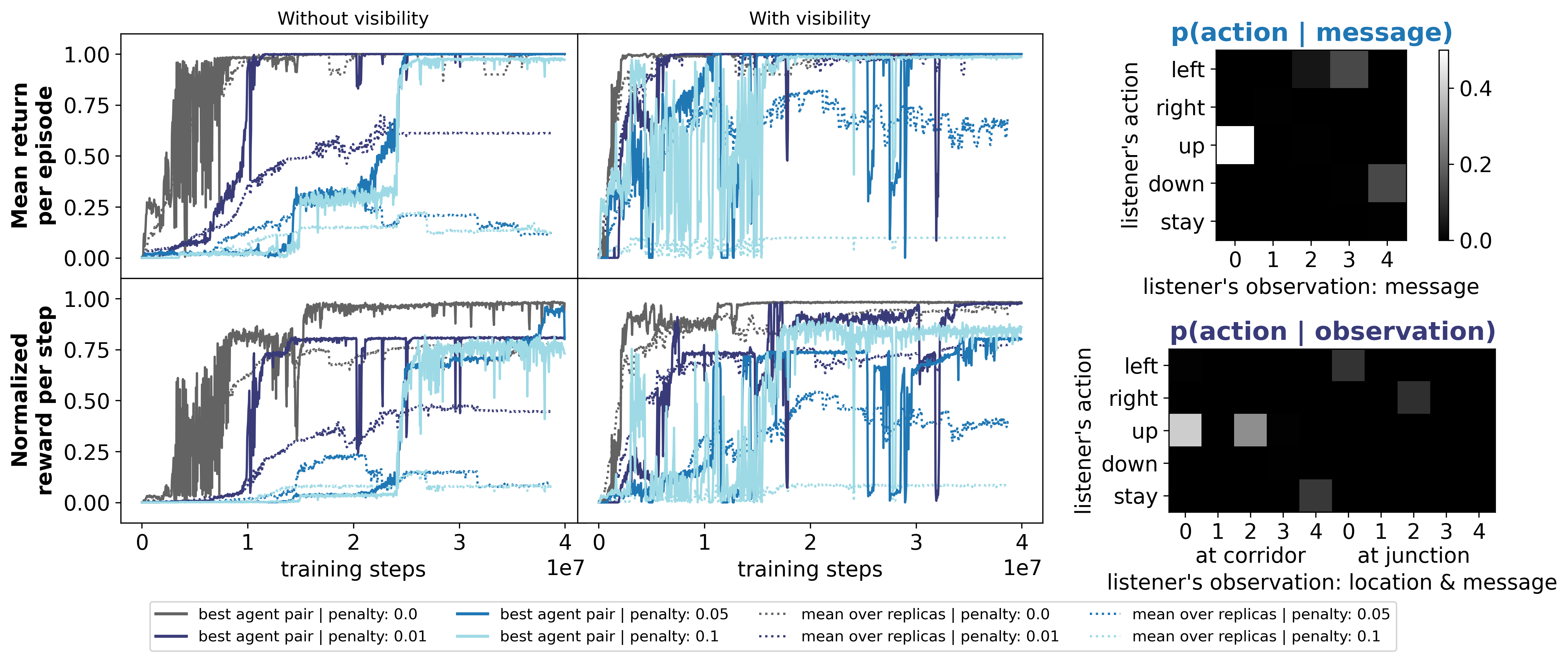}
\includegraphics[width=\textwidth]{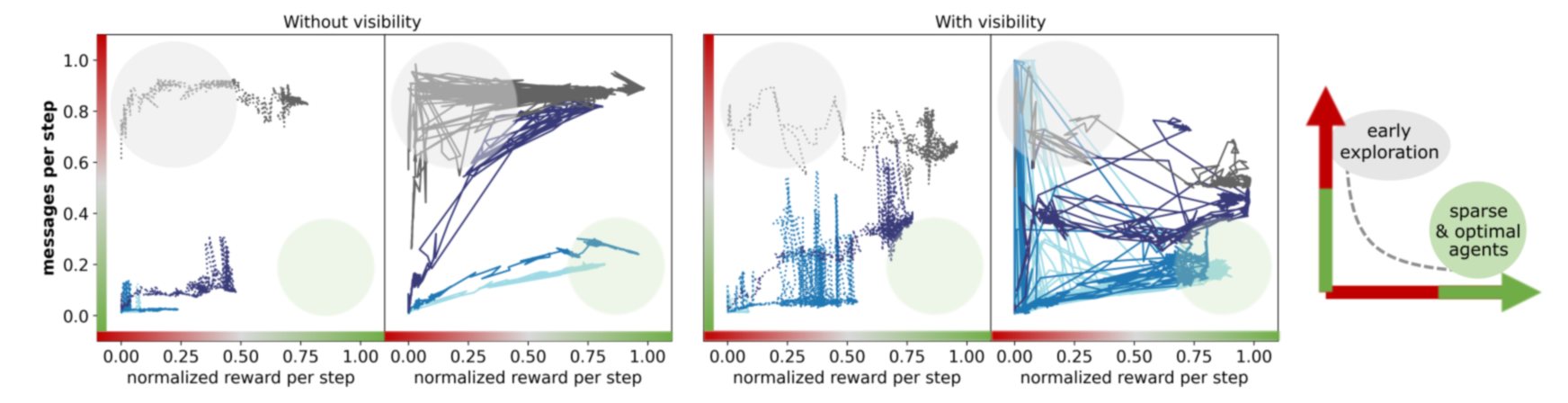}
\end{center}
\caption{\textbf{Communication emergence with a cost on communication effort in the T-maze.} Agents without memory. (top) The additional pressure makes learning more difficult---task performance for blue lines is lower than for the grey lines (agents using cheap talk). (bottom) The immediate cost of non-zero messages stifles early exploration. } 
\label{fig: penalties no memory}
\end{figure}

\textbf{A fixed penalty improves communication sparsity but makes communication emergence more difficult.} We test four fixed penalty values to evaluate the impact of a fixed penalty on communication emergence. In Fig.~\ref{fig: penalties no memory}, we illustrate the learning curves for agents training with a fixed penalty. Note that very few agent pairs learn to reliably solve the task. Only a few agent pairs find a quasi-optimal solution. Secondly, note in the bottom plots in Fig.~\ref{fig: penalties no memory} that when a cost of articulation is introduced, speakers send few non-zero messages at the beginning of training. The decrease in early exploration of a common language likely makes it harder for agents to establish a successful communication protocol and consequently causes difficulty in collaborating to reach the target. 

In summary, when agents converge to a successful communication protocol, the per-message penalty incentivises sparse communication. However, overall convergence rates are low. Out of the three tested penalty values, a fixed penalty $=0.05$ seems to perform best. In Fig.~\ref{fig: pressures comparison} and Fig.~\ref{fig: comparison 3env} for comparison with other pressures, we include agents trained with a fixed penalty $=0.05$.

\textbf{A penalty curriculum does not negatively impact performance but it does not as strongly incentivise conciseness.} To mitigate the issue of stifled early exploration, we implement penalty curricula, in which agents start training with no per-message penalty. Only after they reach a performance threshold (as described in Section~\ref{sec: experimental setup}), the speaker begins to experience a penalty for every non-zero message. As visible in Fig.~\ref{fig: pressures comparison}, this strategy significantly improves convergence rates---agents reliably learn to communicate \textit{on par} with baseline agents. However, communication is not sparse. The agents exchange only slightly fewer messages than baseline agents (refer to the green vs. grey lines in Fig.~\ref{fig: pressures comparison} row 3). Once they find a communication protocol that works, they seem unlikely to change it even with an increasing penalty---the sparsity metric remains relatively constant throughout training. The trend is very similar in the other two mazes (see Fig.~\ref{fig: comparison 3env}). 

All in all, introducing a penalty (fixed or progressive) makes the communication sparser compared to baseline. However, the sparsity of communication is still suboptimal.

\subsection{\textit{Situated} communication}

\textbf{The pressure of time in a multi-step interaction can incentivise sparse communication without compromising communication emergence.} In the last set of experiments, we evaluate the impact of \textit{situated} communication on language emergence. We find that \textit{situated} communication incentivises conciseness without stifling early exploration. Active listeners learn over time that communication comes at a cost, adjusting \textit{when} and \textit{whether} they solicit information, rather than avoiding communication early on (refer to the orange lines in Fig.~\ref{fig: pressures comparison} row 3). After training, agents using \textit{situated} communication achieve highest communication sparsity. Importantly, the sparsity does not result in a loss in task performance---agents with an active listener follow learning patterns similar to baseline agents (refer to Fig.~\ref{fig: pressures comparison} row 1 \& 2).

In a T-maze environment, the active listener can learn to near optimally solicit information, asking \textapprox$2.06$ and \textapprox$9.76$ times per episode given partial and no visibility, respectively. In order to be theoretically optimal under the no visibility condition, an active listener without memory requires $9$ messages per episode ($1$ message per step) and, under partial visibility, $2$ messages per episode ($1$ at each junction). The heatmaps in Fig.~\ref{fig: heatmaps} illustrate example communication protocols of successful agent pairs. Under the no memory and no visibility condition, the listener queries the speaker at each timestep, in line with the optimal strategy. Under the partial visibility condition, information solicitation takes place mostly at the junctions, where the acting agent has a choice between two viable environmental actions.  

Interestingly, when we test \textit{situated} communication between agents with memory, agents continue to ask for information at the junctions when it is immediately actionable, as visible in the top right heatmap in Fig.~\ref{fig: heatmaps}. This is non-obvious---given memory, the active listener could ask for information at any point in the maze. This result suggests that it is easier for agents to succeed at the task when they exchange information when it is immediately actionable. There might be benefit to allowing agents to reason about the timing of communication in multi-step interactions. We show more results that support this thesis in Appendix~\ref{appendix: multistep-vs-upfront}.

Lastly, we test the robustness of \textit{situated} communication across two other environments: dead ends and four rooms. As visible in Fig.~\ref{fig: comparison 3env}, across all three environments, active listeners learn to reliably solve the task---they achieve a mean return per episode equal to $1$---while requiring the least amount of communication---their sparsity metric is closest to $0$. In some mazes, active listeners converge to solutions that are slightly suboptimal compared to agents using other modes of communication, meaning that on average they take more steps to reach the target. Future work will investigate further how agents using \textit{situated} communication can learn to solve the task optimally. A symmetrical implementation of \textit{situated} communication, where both agents can act in the environment, might offer a robust solution. 

\section{Conclusions \& Discussion}
We find that giving the listener agency to choose \textit{whether} to communicate enables agents to learn how to concisely exchange task-relevant information. By \textit{situating} the communication in the task, we allow the functional pressure of time to shape the emergent communication protocol to be sparse, in line with the Gricean maxim of quantity. A cost of articulation, implemented in the form of a per-message penalty, is not as effective. 

As common in related work and described in Section~\ref{sec: communication modes}, the current implementation of communication is asymmetrical---only the speaker has privileged information and only the listener can take non-communicative actions in the environment. Our ongoing work will expand this implementation and situate both the speaker and listener in the environment, enabling both agents to communicate and take non-communicative actions in the maze-bazed world. Moreover, we would like to explore more complex cooperative tasks, such as those in the Overcooked game environment~\cite{carroll2019utility,knott2021evaluating}. One of the objectives of achieving concise communication protocols is to design RL agents that can effectively communicate with people. In the work's current format, we anticipate significant societal benefit without negative societal impacts.

\appendix
\section{}
 
\subsection{Implementation details} 
\label{Implementation}
\begin{figure}[t]
\begin{center}
\includegraphics[width=\textwidth]{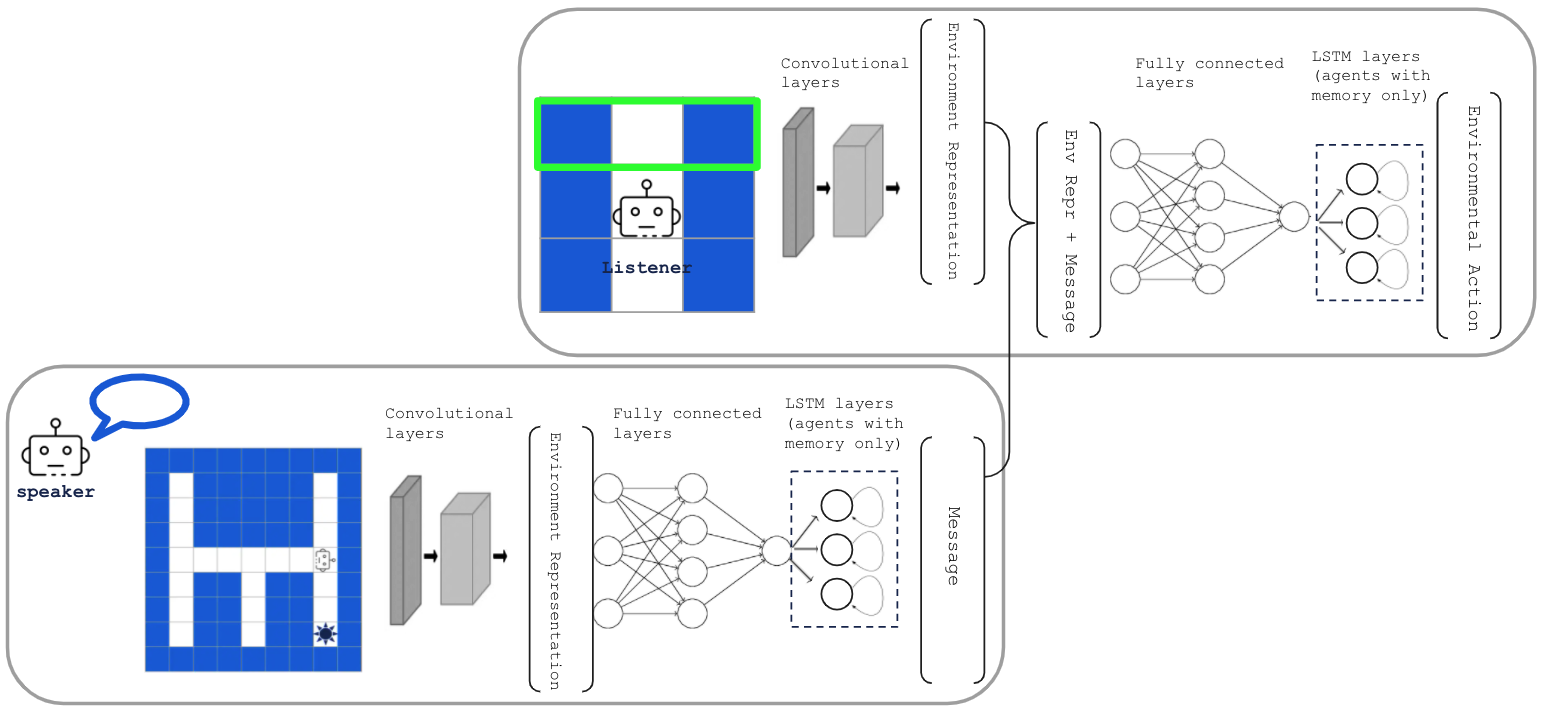}
\end{center}
\caption{\textbf{Agents' architecture and mechanism of communication.}} 
\label{fig: agent_architecture}
\end{figure}

Agent architecture is visualized in Fig.~\ref{fig: agent_architecture}. All models were implemented in Python using JAX~\cite{bradbury2020jax} and Haiku~\cite{haiku2020github}. For training, we used GPUs V100 and P100.

\begin{figure*}[t]
\begin{center}
\includegraphics[width=\textwidth]{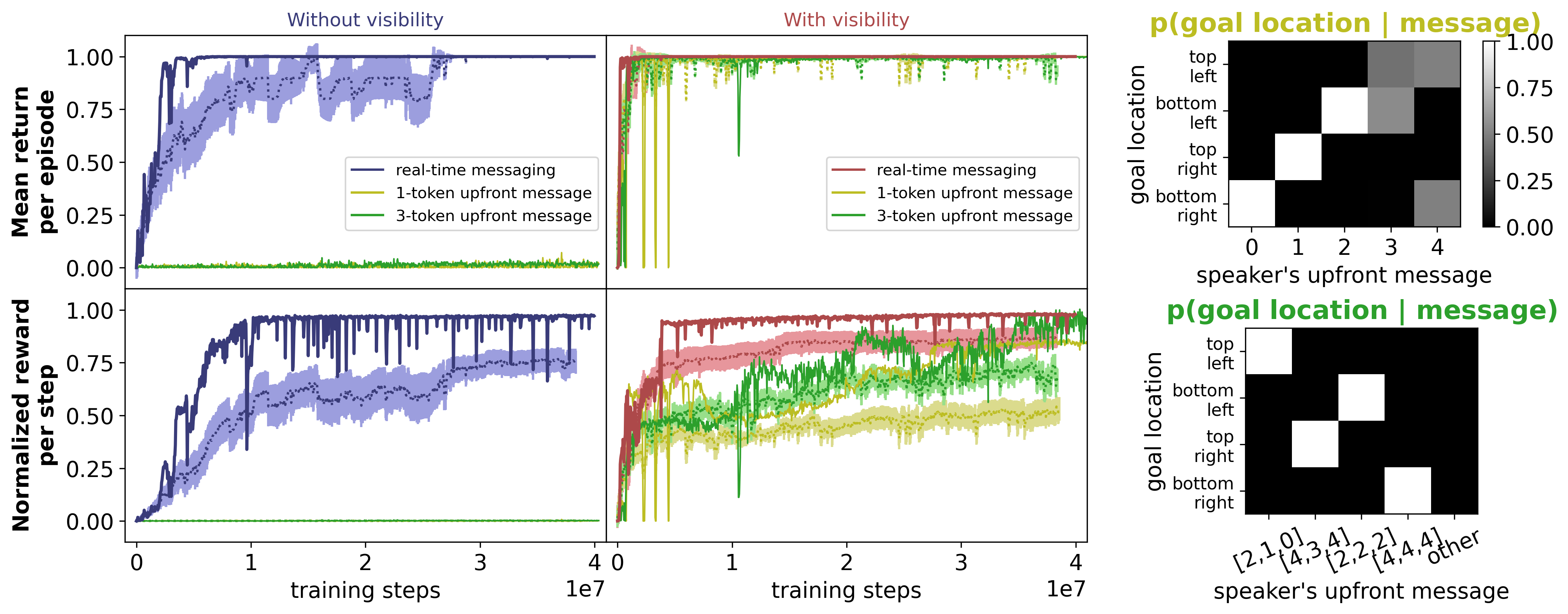}
\end{center}
\caption{\textbf{Comparison of upfront and real-time messaging.} Agents have memory. Real-time messaging improves convergence on a successful communication protocol. With upfront messaging, agents learn to solve the task before episode timeout when the listener has partial visibility. However, convergence is slow and agents are unlikely to solve the task in the optimal number of steps. } 
\label{fig: upfront messaging}
\end{figure*}

\subsection{Experimental comparison of upfront vs. real-time messaging} 
\label{appendix: multistep-vs-upfront}

\textbf{Real-time communication improves language emergence compared to upfront messaging.} As an additional experiment, we compare real-time communication (multi-step cheap talk) with upfront messaging. In upfront messaging, the speaker generates a 1-, 2-, or 3-token message at the beginning of each episode and that message gets broadcasted to the listener at each timestep throughout the episode. It is a form of \textit{unsituated} communication, where the speaker can only generate one message at the beginning of an episode. In both scenarios, the theoretical capacity of the communication channel allows the agents to communicate the necessary information, whether the agents choose to communicate directions, e.g., `turn right', or a goal address, e.g., `top left corner'. With upfront messages of length $1$, $2$, and $3$, the speaker has $5$, $25$, or $125$ unique messages available for communication, respectively.  

In Fig.~\ref{fig: upfront messaging}, we plot the best agent pairs as well as the mean over replicas with the same hyperparameters as the best agent pair. With both real-time and upfront messaging, agents succeed in establishing a successful communication protocol when the listener has partial visibility---they converge to a mean return of $1$ per episode. With no visibility for the acting agent, agent pairs with upfront messaging do not succeed at solving the task. Moreover, the real-time agents are more likely to converge to an optimal solution, being able to solve the T-maze task in $9$ moves. With $1$ upfront token, even the best agents learn to at-best solve the task in $12$ steps. These agents seem to reliably learn unique messages to encode the action required at the first turn or the right/left part of the address, but they do not establish a unique encoding for the top/bottom portion of the address, as visible in the top heatmap in Fig.~\ref{fig: upfront messaging}. With $3$ upfront tokens, the best agent pair agrees on $4$ distinct symbols to encode the $4$ possible goal locations. However, convergence is slow and on average agent pairs perform less optimally than under the real-time communication paradigm. We hypothesize that there are benefits to allowing communication to emerge from multi-step interactions. Our findings suggest that it is easier for agents to learn to communicate if they can exchange information when it is immediately actionable.

\begin{figure}[t]
\begin{center}
\includegraphics[width=\textwidth]{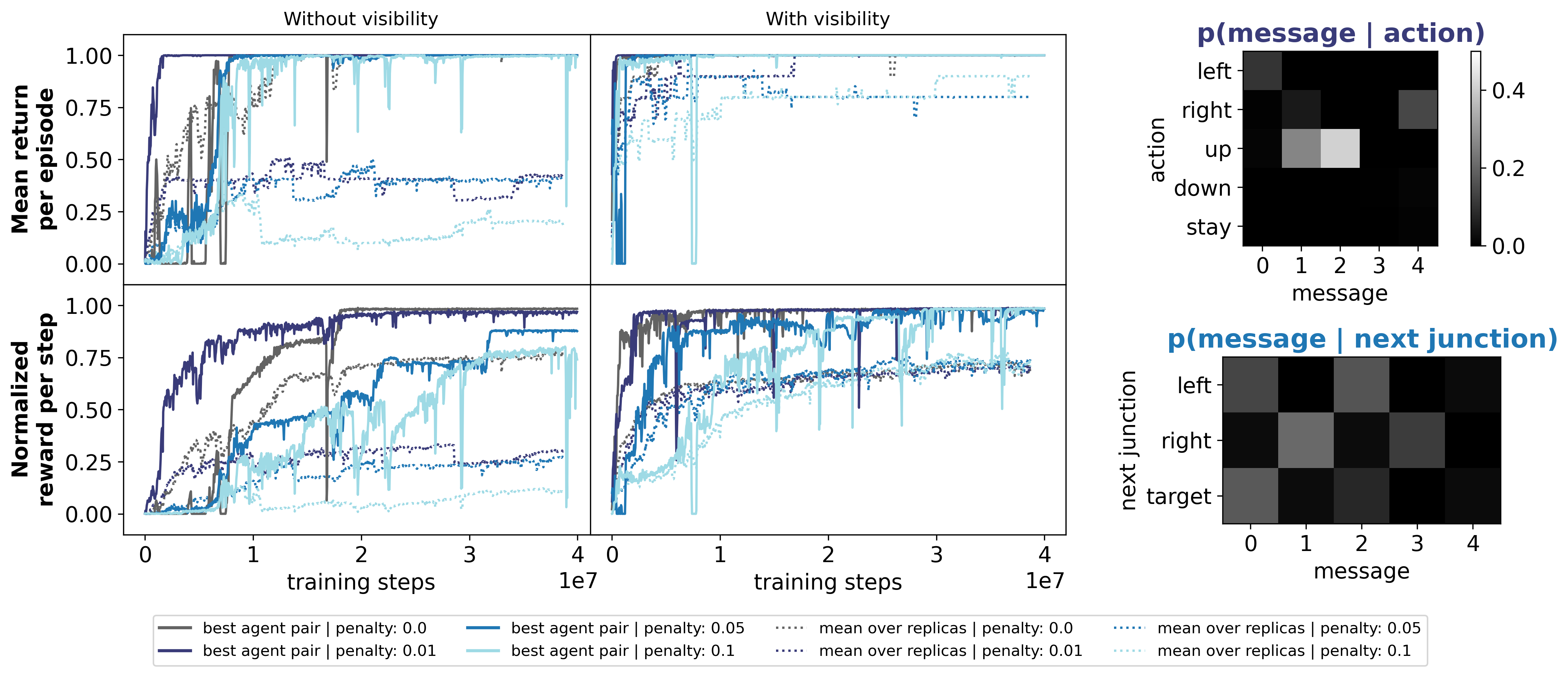}
\includegraphics[width=\textwidth]{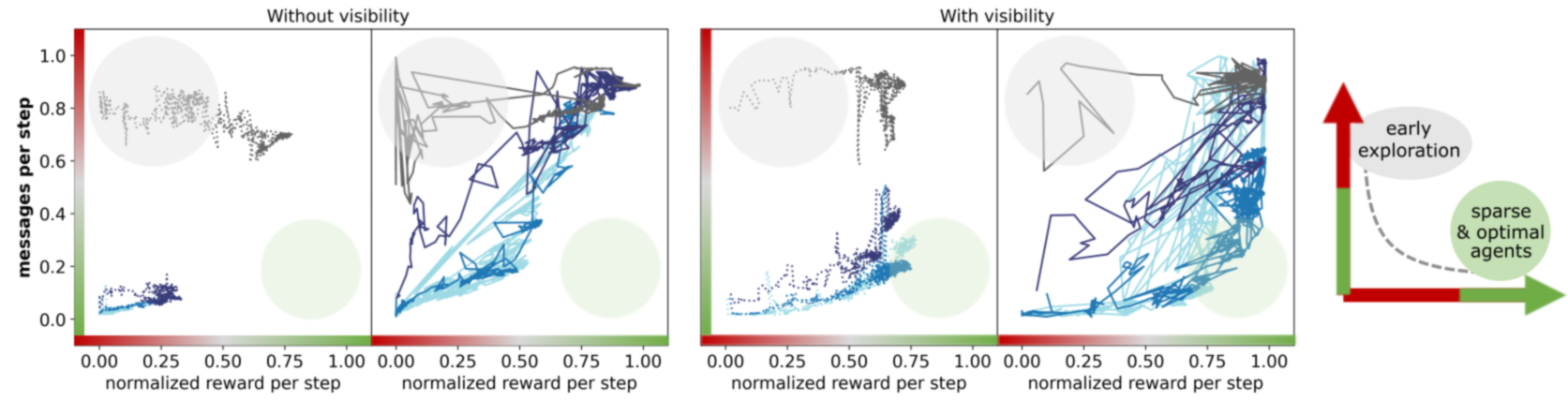}
\end{center}
\caption{\textbf{Communication emergence with a cost on communication effort.} Agents with memory; speaker experiences a per-message penalty. With memory, overall performance improves---more agent pairs converge to an optimal solution. However, relative to baseline, agents with a message penalty have more difficulty learning to jointly solve the task. In the heatmaps on the right, we illustrate the communication protocol of the best agent pairs. Note that the penalty is largely ineffective---the agents send many non-zero messages per episode.} 
\label{fig: penalties & memory}
\end{figure}

\begin{figure}[ht]
\begin{center}
\includegraphics[width=\textwidth]{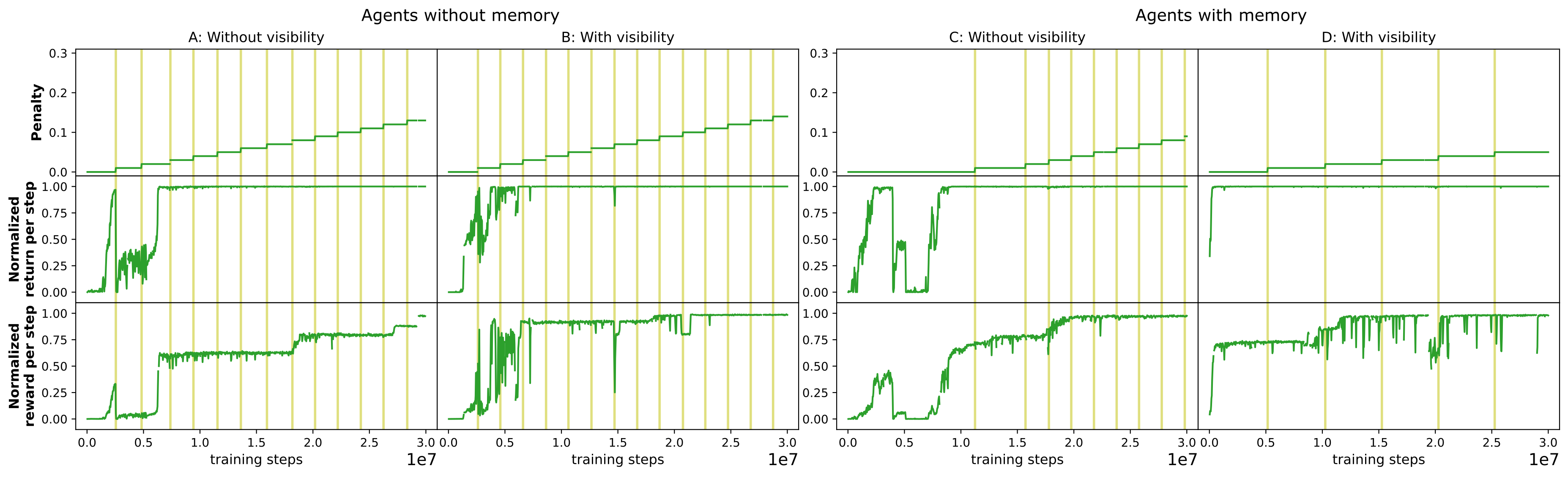}
\includegraphics[width=\textwidth]{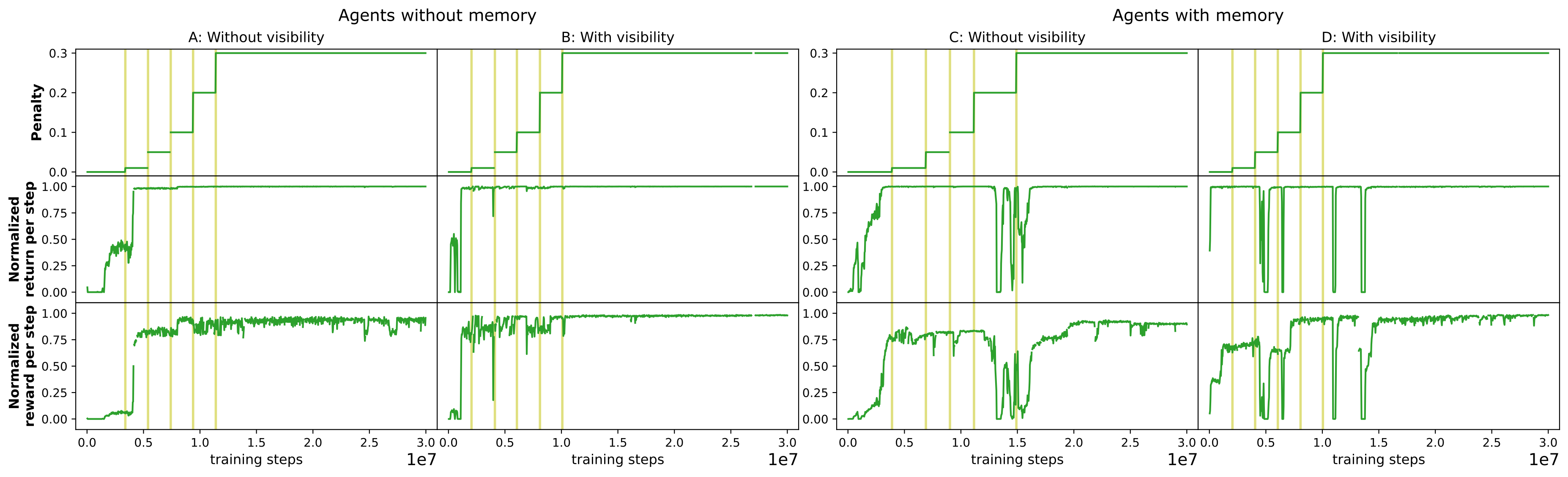}
\end{center}
\caption{\textbf{Best agent pairs training with a penalty curriculum.} (top) Curriculum with mapping $m_{p1}$ (bottom) curriculum with mapping $m_{p2}$.} 
\label{fig: curriculum}
\end{figure}

\subsection{Additional results for agents with a per-message penalty} 

In the main paper, we present results for experiments on agents without memory. In Fig.~\ref{fig: penalties & memory}, we present the same experiment on agents with memory. Overall, agents with memory perform better than agents without memory. However, the trends remain the same---the penalty negatively impacts convergence and the best-performing agents that find optimal task solutions do not exhibit sparse communication protocols. 

In Fig.~\ref{fig: curriculum}, we visualize the learning curves for the two curricula in parallel with the agents' progression through the curriculum stages. The curriculum with mapping $m_{p1}$ achieves better overall performance and is included in the comparison plots in the main paper (Fig.~\ref{fig: pressures comparison} and Fig.~\ref{fig: comparison 3env}). 

\subsection{Additional results for agents using \textit{situated} communication} 

\begin{figure*}[t]
\begin{center}
\includegraphics[width=\textwidth]{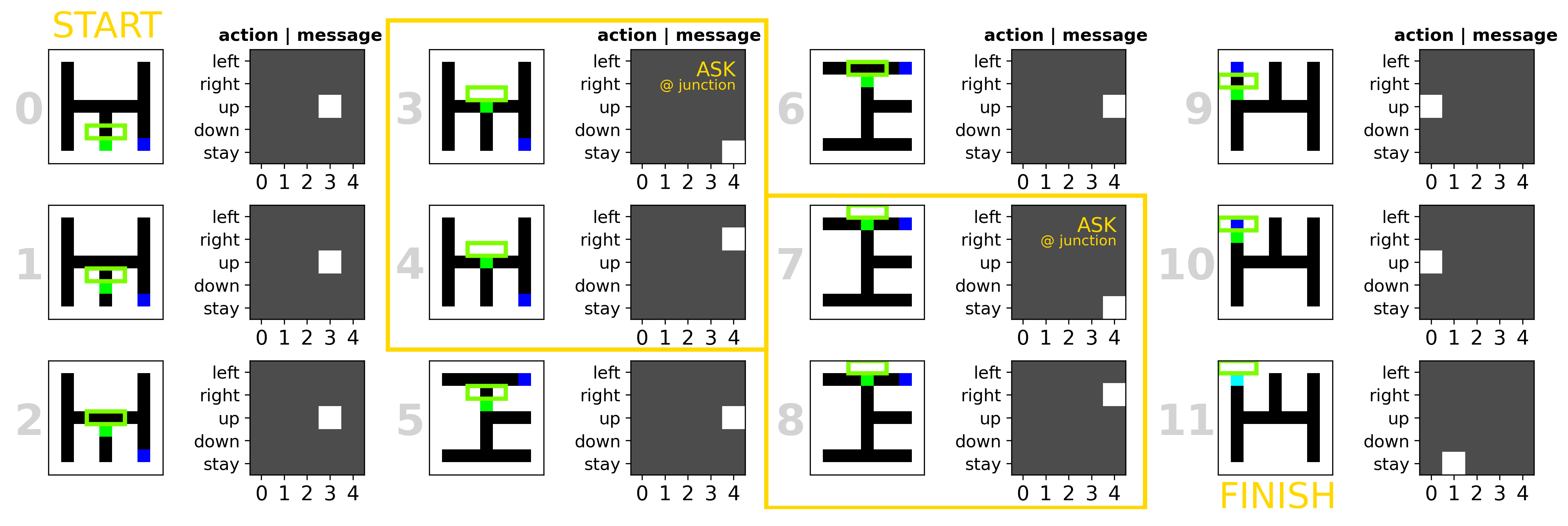}
\end{center}
\caption{\textbf{Walk-through of an example episode with \textit{situated} communication.} The listener learns to solve the task optimally, deciding to stay and ask for information when at a junction (twice during the episode). At each of the $11$ timesteps, we visualize (left) the speaker's view of the board with an overlaid green box indicating the listener's view, and (right) the speaker's message and listener's action at that step. }
\label{fig: active-listener-walkthrough}
\end{figure*}

Fig.~\ref{fig: active-listener-walkthrough} shows a step-by-step example episode for a listener with partial visibility. The gridworld colors correspond to the color encoding included in the agents' observation vectors during the task. 

\subsection{Additional results on the impact of pressures on communication conciseness}

\begin{figure*}[h]
\begin{center}
\includegraphics[width=\textwidth]{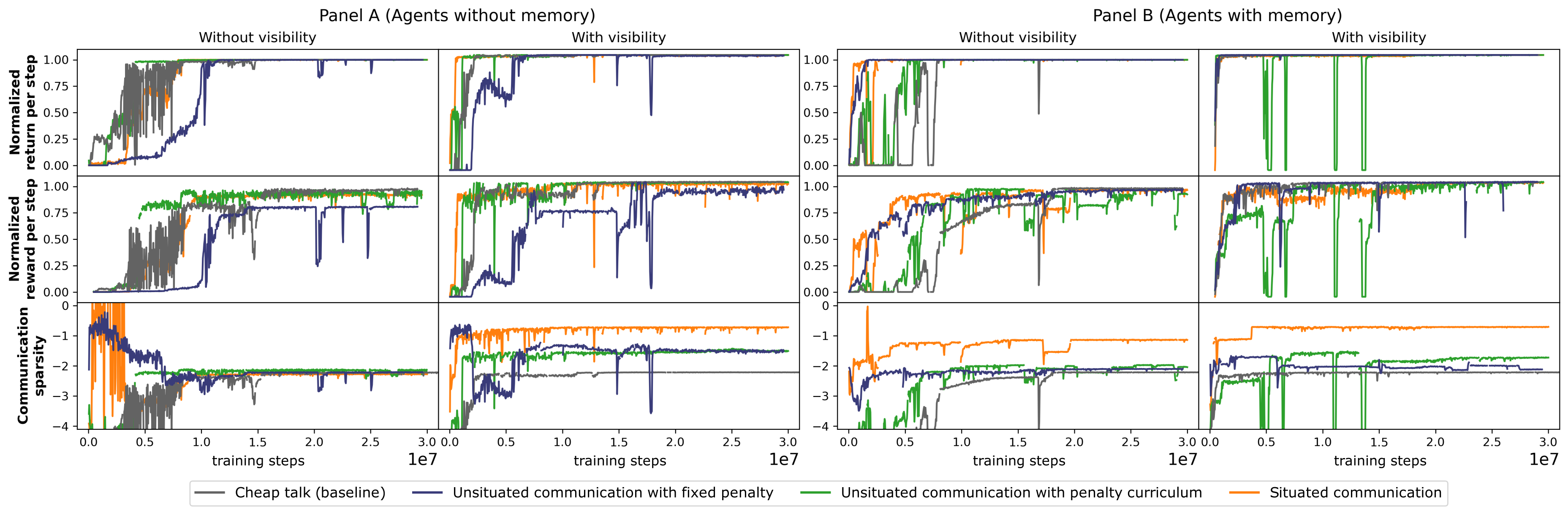}
\end{center}
\caption{
\textbf{Comparison of pressures for avoiding over-communication (best agent pairs) in the T-maze.}} 
\label{fig: comparison-tmaze-best}
\end{figure*}

\begin{figure*}[h]
\begin{center}
\includegraphics[width=\textwidth]{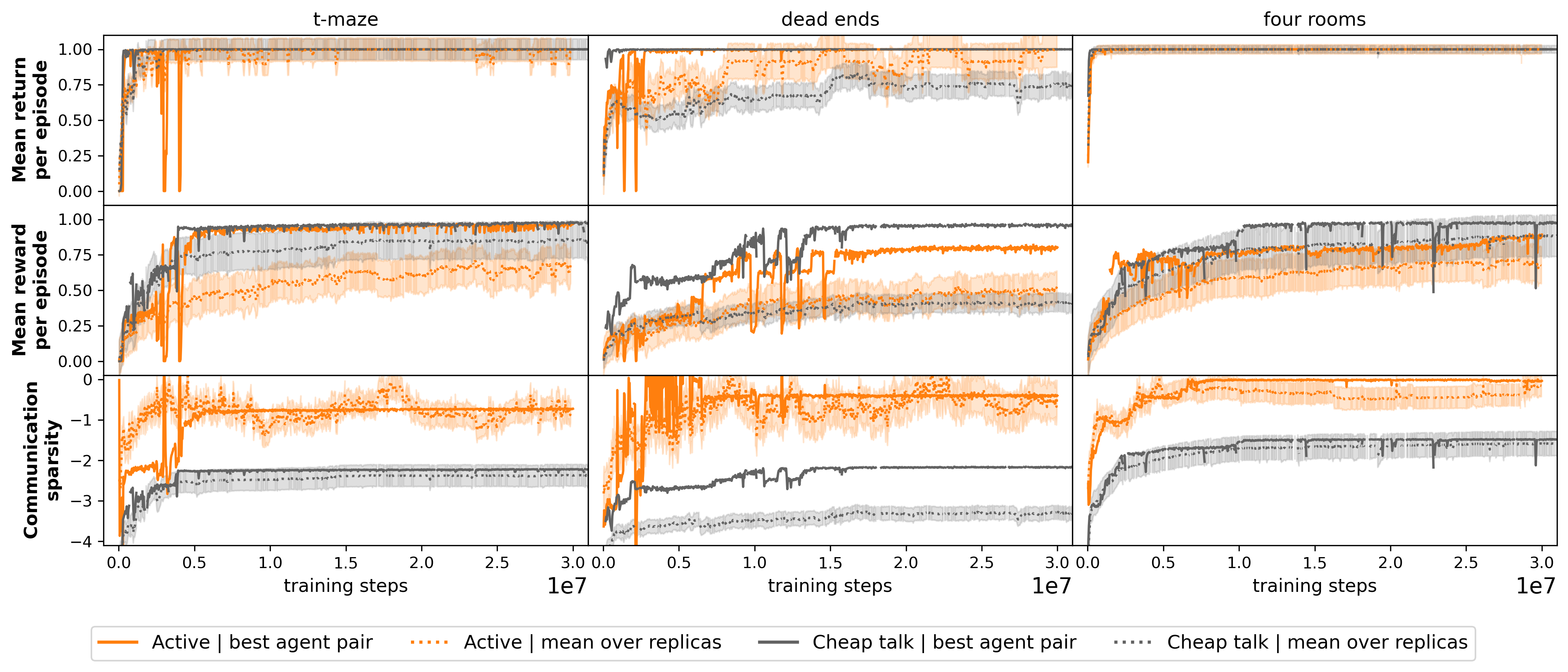}
\includegraphics[width=\textwidth]{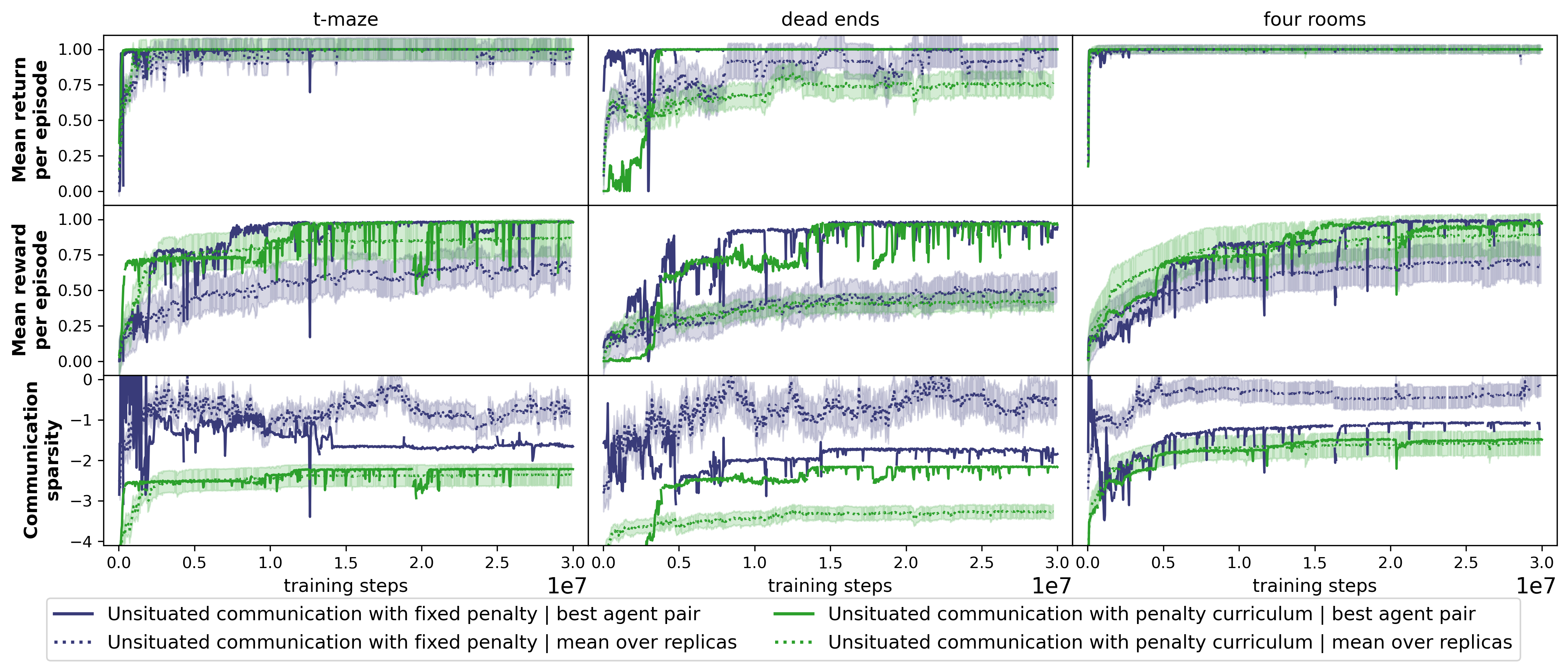}
\end{center}
\caption{\textbf{Agents in 3 environments} (top) using \textit{situated} communication compared to cheap talk, (bottom) using \textit{unsituated} communication with a fixed penalty and penalty curriculum.} 
\label{fig: 3env-additional-results}
\end{figure*}

In the main text, we plot mean learning curves selected based on the highest mean normalized reward per step for a given set of hyperparameters. Here, we include results from the same experiments as in the main text, but instead we plot learning curves for the best agent pair. Fig.~\ref{fig: comparison-tmaze-best} includes the same comparison as Fig.~\ref{fig: pressures comparison} in the main text; Fig.~\ref{fig: 3env-additional-results} includes the same comparison as Fig.~\ref{fig: comparison 3env}. In Fig.~\ref{fig: 3env-additional-results}, we present the best performing agent pairs from our hyperparameter sweep as well as the mean over the $10$ replicas with the same hyperparameters as the best performing pair. The best performing agent pairs are selected based on the metric of solution optimality (the normalized reward per step). Overall, the trends of the best performing agent pairs are consistent with the mean trends described in the paper. 

\bibliography{references}
\bibliographystyle{theapa}

\end{document}